\documentclass{article}[12pt]
\usepackage{setspace}
\usepackage{psfig}
\usepackage[super]{natbib}

\begin{document}

\begin{center}
{\Large \bf 
\noindent
{An ultra-luminous quasar with a twelve-billion-solar-mass black hole at redshift 6.30}}
\end{center}
\vspace{4mm}

%\singlespacing
\begin{center}
\noindent
{\parbox{\textwidth}{\raggedright Xue-Bing Wu$^{1,2}$,
%\thanks{E-mail: \texttt{wuxb@pku.edu.cn}}
Feige Wang$^{1,2}$, 
Xiaohui Fan$^{2,3}$,
Weimin Yi$^{4,5,6}$,
Wenwen Zuo$^{7}$,
Fuyan Bian$^{8}$,
Linhua Jiang$^{2}$,
Ian D. McGreer$^{3}$, 
Ran Wang$^{2}$, 
Jinyi Yang$^{1,2}$,
Qian Yang$^{1,2}$,
David Thompson$^{9}$
and Yuri Beletsky$^{10}$
}\vspace{0.4cm}\\
\parbox{\textwidth}{\raggedright $^{1}$Department of Astronomy, School of Physics, Peking University, Beijing 100871, China\\
$^{2}$Kavli Institute for Astronomy and Astrophysics, Peking University, Beijing 100871, China\\
$^{3}$Steward Observatory, University of Arizona, Tucson, AZ 85721-0065, USA\\
$^{4}$Yunnan Observatories, Chinese Academy of Sciences, Kunming 650011, China\\
$^{5}$University of Chinese Academy of Sciences, Beijing 100049, China\\
$^{6}$Key Laboratory for the Structure and Evolution of Celestial Objects, Chinese Academy of Sciences, Kunming 650011, China\\
$^{7}$Shanghai Astronomical Observatory, Chinese Academy of Sciences, Shanghai 200030, China\\
$^{8}$Mount Stromlo Observatory, Research School of Astronomy and Astrophysics, Australian National University,
Weston Creek, ACT 2611, Australia\\
$^{9}$Large Binocular Telescope Observatory, University of Arizona, Tucson, AZ 85721, USA\\
$^{10}$Las Campanas Observatory, Carnegie Institution of Washington, Colina el Pino, Casilla 601, La Serena, Chile
}}
\vspace{5mm}
\end{center}

%\doublespacing

{\bf
\noindent
So far, roughly 40 quasars with redshifts greater than $z=6$ have been 
discovered\cite{fan03,jiang07,willott07,jiang08,willott10b,mortlock11,venemans13,banados14}.
Each quasar contains a black hole with a mass of about one billion solar masses
( $10^9 M_\odot$) \cite{jiang07,mortlock11,venemans13,willott03,kurk07,willott10,derosa11,derosa14}. 
The existence of such black holes when the Universe was less than 1 billion years old presents substantial challenges to 
 theories of the formation and growth of black holes and the coevolution of black holes and galaxies\cite{volonteri12}.
Here we report the discovery of an ultra-luminous quasar, 
SDSS J010013.02+280225.8, at redshift $z=6.30$. It has an  optical and near-infrared luminosity a few times 
greater than those of previously known $z>6$ quasars. 
On the basis of the deep
absorption trough\cite{gunn65} on the blue side of the Ly $\alpha$ emission line in the spectrum, we estimate the proper size of the ionized proximity zone associated with the quasar to be
26 million light years, larger than found with other $z>6.1$ quasars with 
lower luminosities\cite{fan06}.
We estimate (on the basis of a near-infrared spectrum) that the black hole has a mass of $\sim 1.2 \times 10^{10} M_\odot$,
which is consistent with the $1.3 \times 10^{10} M_\odot$ derived by assuming an Eddington-limited accretion rate. 
}

High redshift quasars have been efficiently selected using a combination of  optical and near-infrared 
colours\cite{willott07,jiang08}.  
We have carried out a systematic survey of quasars at $z>5$ using photometry 
from the Sloan Digital Sky Survey (SDSS)\cite{york00}, the two Micron All Sky Survey (2MASS)\cite{skrutskie06} 
and the Wide-field Infrared Survey Explorer (WISE)\cite{wright10}, resulting in the discovery of a significant population of luminous high redshift quasars.   
SDSS J010013.02+280225.8 (hereafter J0100+2802), was selected as a high-redshift quasar candidate due to its red 
optical colour (with SDSS AB magnitudes  
$i_{AB}=20.84\pm0.06$ and $z_{AB}=18.33\pm0.03$) and a photometric redshift of $z\simeq 6.3$. It has bright 
detections in the 2MASS $J$, $H$ and $K_s$ bands with Vega magnitudes of 17.00$\pm$0.20, 15.98$\pm$0.19, 
and 15.20$\pm$0.16, respectively; it is also strongly detected in WISE, with  Vega magnitudes in W1 to W4 bands of 14.45$\pm$0.03, 13.63$\pm$0.03, 
11.71$\pm$0.21, and 8.98$\pm$0.44 (see Extended Data Figs 1 and 2 for images in different bands). 
Its colour in the two bluest WISE bands, W1-W2, clearly separates it from the bulk of stars 
in our Galaxy \cite{wu12}.
The object was within the SDSS-III imaging area. It is close to the colour selection boundary of SDSS $z\sim6$ quasars\cite{fan03}, 
but was assigned to low priority earlier because of its relatively red $z_{AB}-J$ colour and its bright apparent magnitudes. 
It is undetected in both radio and X-ray bands by the wide-area, shallow survey instruments.

Initial optical spectroscopy on J0100+2802 was carried out on 29 December, 2013 with 
the Lijiang 2.4-m telescope in China. 
The low resolution spectrum
clearly shows a sharp break at about 8,800$\AA$,  
consistent with a quasar at a redshift beyond 6.2. 
Two subsequent
optical spectroscopic observations were conducted on 9 and 24 January 2014 respectively with the 6.5-m Multiple 
Mirror Telescope (MMT) and
the twin 8.4-m mirror Large Binocular Telescope (LBT) in the USA.
The Ly$\alpha$ (Ly$\alpha$) line shown in the 
spectra confirms that J0100+2802 is a quasar 
at redshift of 6.30$\pm$0.01 (see Fig. 1 and Methods for details).  

We use the multiwavelength photometry to estimate the optical luminosity 
at rest-frame 3,000$\AA$($L_{3000}$), which is consistent with that obtained from K-band spectroscopy (see below). 
The latter gives a more reliable value of 3.15$\pm0.47 \times 10^{47}~{\rm ergs~s}^{-1}$, adopting a $\Lambda$CDM cosmology with 
Hubble constant $H_0=70 {\rm km~s}^{-1}~Mpc^{-1}$, matter density parameters $\Omega_M=0.30$ and dark energy density parameter $\Omega_{\Lambda}=0.7$.
Assuming
an empirical conversion factor from the luminosity at 3,000$\AA$ to the bolometric luminosity \cite{shen11}, this
gives $L_{\rm bol}=5.15 \times L_{3,000}= 1.62\times 10^{48} {\rm ergs~s}^{-1} = 4.29 \times 10^{14} L_{\odot}$ 
(where $L_{\odot}$ is the solar luminosity).
We obtain a similar result when estimating the bolometric luminosity from the Galactic extinction corrected absolute magnitude at
rest-frame 1450$\AA$, which is $M_{1450,AB}=-29.26\pm0.20$.
The luminosity of this quasar is roughly 4 times greater than that of the luminous $z=6.42$ quasar\cite{fan03} SDSS J1148+5251, and 
7 times greater than that of the most distant known quasar\cite{mortlock11} ULAS J1120+0641 ($z=7.085$); 
it is the most luminous
quasar known at $z>6$ (see Extended Data Fig. 3).

The equivalent width (EW) of the Ly$\alpha$+ N V emission lines as measured from the LBT spectrum is roughly 10 $\AA$, 
suggesting that J0100+2802 is probably a weak-line quasar (WLQ) \cite{fan99}. 
WLQ fraction is higher among the $z\simeq6$ quasars compared to that at lower redshift\cite{banados14}, and
a high detection rate of strong 
millimeter dust continuum in z$\sim$6 WLQs points to active star formation in these
objects\cite{wang08}.
Given its extreme luminosity, J0100+2802 will be helpful in the study of the evolutionary 
stage of WLQs by future 
(sub)millimeter observations,
though the origin of the weak UV emission line feature of WLQs is still uncertain.

The LBT spectrum of J0100+2802 (Fig. 1) exhibits a deep Gunn-Peterson absorption trough\cite{gunn65}  
blueward of the Ly$\alpha$ emission. The transmission spectrum (assuming an intrinsic power-law continuum\cite{van01} 
of $F_{\lambda} \propto \lambda ^{-1.5}$) is shown in Fig. 2. Complete Gunn-Peterson absorption 
can also be seen in the Ly$\alpha$, Ly$\beta$ and Ly$\gamma$ transitions. Statistically significant 
transmission peaks are detected at $z=5.99$ in both the Ly$\alpha$ and Ly$\beta$ troughs, and an additional 
transmission peak is detected at $z=5.84$ in the Ly$\beta$ trough. The 2$\sigma$ lower limit on the 
Ly$\alpha$ Gunn-Peterson optical depth ($\tau_{\alpha}$) at $z=6.00 \mbox{--} 6.15$ is $\tau_{\alpha}>5.5$ and the 
2$\sigma$ lower limit for Ly$\beta$ is $\tau_{\beta}>6$, corresponding to an equivalent $\tau_{\alpha} >13.5$, 
following the conversion in literature\cite{fan06}.
The characteristics of the intergalactic medium (IGM) transmission along the line of sight of J0100+2802, including the 
deep Ly$\alpha$ and Ly$\beta$ troughs, and the narrow, unresolved transmission peaks, are similar to those 
observed in SDSS J1148+5251, and are consistent with the rapid increase in the 
IGM neutral fraction at $z>5.5$ observed in a large sample of SDSS quasars\cite{fan06}. 
The size evolution of quasar proximity zone, which is highly ionized by quasar UV photons,  
can also be used to constrain the IGM neutral fraction. The proximity zone size is 
defined as the point where the transmitted flux first drops by a significant amount to below 10\% (ignore small absorption leaks) of the quasar extrapolated continuum emission after the spectrum is smoothed to a resolution of 20$\AA$ \cite{fan06}. As shown in Fig. 2, J0100+2802 
has a much larger proper proximity zone ($7.9\pm0.8$ Mpc; 1 Mpc is about 3.26 million light years) than that of other
SDSS quasars\cite{fan06,carilli10} at $z>6.1$; its large proximity zone size is expected from the higher level of  
photo-ionization dominated by quasar radiation.

We obtained the near-infrared J,H,K-band spectra with Gemini and Magellan telescopes on 6 August and 7 October 2014, respectively (see Methods for details). 
Figure 3 shows the combined optical/near-infrared spectrum of J0100+2802 and the result of fitting  the Mg II emission line.
The Mg II Full Width at Half Maximum (FWHM) is $5,130\pm150~{\rm km~s}^{-1}$, 
and the continuum luminosity at the rest-frame wavelength of 
3,000$\AA$  is $3.15\pm0.47 \times 10^{47} {\rm ergs~s}^{-1}$. After applying a virial black hole
 mass estimator based on the Mg II line\cite{MD04}, we estimate its black hole mass to be $(1.24\pm 0.19) 
\times 10^{10} M_{\odot}$ ($M_{\odot}$. The uncertainty of black hole mass does not include the systematic uncertainty of
virial black hole mass estimation, which could be up to a factor of three\cite{peterson04}. 
Assuming that this quasar is accreting at Eddington accretion rate and the bolometric luminosity is
close to the Eddington Luminosity ($L_{Edd}=1.3\times10^{38}(M/M_{\odot})$), 
similar to other $z>6$ quasars\cite{willott10}, 
leads to a  black hole mass of $1.3 
\times 10^{10} M_{\odot}$ for J0100+2802. 
Therefore, our observations strongly indicate that J0100+2802 harbors a black hole of mass about $1.2 
\times 10^{10} M_{\odot}$, the first such system known at $z>6$, though  black holes of such a size have been found in local giant elliptical 
galaxies\cite{mcconnell11} and low-redshift quasars\cite{shen11}.

Although gravitational lensing is a possible explanation for its high luminosity, we do not expect a large lensing magnification.
An LBT K-band image with seeing of 0.4$^{\prime\prime}$ shows a morphology fully consistent with a single point source (see Extended Data Fig. 2); 
and the large quasar proximity zone size further supports a high UV luminosity consistent with the expected photoionization scaling \cite{haiman05}. However, absorption features at different redshift have been identified from its near-infrared spectroscopy (see Extended Data Fig. 4), implying the existence of abundant intervening materials along the line of sight.

J0100+2802 is the only quasar with a bolometric luminosity higher than $10^{48}ergs~s^{-1}$ and a black hole
mass larger than $5 \times 10^9 M_{\odot}$ at $z\geq 6$. It is also close to being the most luminous quasar
with the most massive black hole at any redshift (Fig. 4).
The discovery of this ultra-luminous quasar with the entire SDSS footprint ($\sim13,000 degrees^2$) is 
broadly consistent with the extrapolation of SDSS $z\simeq6$ quasar luminosity function\cite{fan06}. 
The number density of such 
objects would set strong constraints on the early growth of supermassive black hole and the evolution of high-redshift quasar 
black-hole mass function\cite{willott10b,willott10}.
In addition to ULAS J1120+0641 with a $2 \times 10^9 M_{\odot}$ black hole\cite{mortlock11,derosa14} at $z=7.085$ and a recently discovered $z=6.889$ quasar witha black hole of $2.1 \times 10^9 M_{\odot}$\cite{derosa14},
J0100+2802 with a  $1.2 \times 10^{10} M_{\odot}$ black hole at $z=6.30$ presents the next most 
significant challenge
to the Eddington-limited growth of black holes in the early Universe\cite{willott10,volonteri12}. 
Its existence 
also strengthens the claim that supermassive black holes in the early Universe probably grew much more quickly than 
their host galaxies, as argued from a molecular gas study of $z\simeq6$ quasars\cite{wang10},
Therefore,  
as the most luminous quasar known to date at $z>6$, J0100+2802 will be a unique resource for the future
study of the mass 
assembly and galaxy formation around the most massive black holes at the end of the epoch of cosmic reionization\cite{fan06}.

\smallskip

\noindent
{\bf Acknowledgments}
X.-B. W thanks the NSFC (grants nos.11033001 and 11373008),  the Strategic Priority Research 
Program "The Emergence of Cosmological Structures" of the Chinese Academy of Sciences (grant no. XDB09000000), and the National 
Key Basic Research Program of China (grant no. 2014CB845700) for support.
X. F., R. W. and I.D. M. thank  the US NSF (grants nos AST 08-06861
and AST 11-07682) for support. R.W. thanks the NSFC (grant no. 11443002) for support. We acknowledge the support of the staffs of the Lijiang 2.4-m telescope. 
Funding for the telescope has been provided by Chinese Academy of Sciences and the People's Government of Yunnan 
Province. This research uses data obtained through the Telescope Access Program (TAP), which has
been funded by the Strategic Priority Research Program ‘The Emergence of
Cosmological Structures’ (grant no. XDB09000000), National Astronomical
Observatories,Chinese Academy ofSciences, and the SpecialFundforAstronomy from
the Ministry of Finance of China. We thank D. Osip for helps in doing Magellan/FIRE
spectroscopy and Y.-L. Ai, L. C. Ho, Y. Shen and J.-G. Wang for valuable suggestions on data analyses.
We acknowledge the use of SDSS, 2MASS, WISE data and of the MMT, LBT, Gemini and Magellan telescopes;  detailed acknowledgments of these facilities
can be found in the Supplementary Information.\\
\noindent{\bf Author Contributions}
X.-B. W, F. W. and X. F. planned the study, and wrote
the draft version of the paper. All other coauthors  contributed
extensively and equally to the observations, data analyses and writing of the manuscript .

\noindent{\bf Author Information}
 Reprints and permissions information is available at
www.nature.com/reprints. The authors declare no competing financial interests.
Readers are welcome to comment on the online version of the paper.
Correspondence and requests for materials should be addressed to
X.-B.W. (wuxb@pku.edu.cn).
\newpage
\noindent{\bf Figure legends}\\
{\bf Figure 1. The optical spectra of J0100+2802}. From top to bottom, the spectra taken with the Lijiang 2.4-m telescope, MMT and LBT 
(in red, blue and black colours), respectively. For clarity, two spectra are offseted upward by one
and two vertical units. Although the spectral
resolution varies from very low to medium ones, in all spectra the 
Ly$\alpha$ emission line, with a rest-frame wavelength of 1,216$\AA$, is redshifted to around 8,900$\AA$, giving
a redshift of 6.30. J0100+2802 is a weak-line quasar with continuum luminosity about four times higher than that of SDSS J1148+5251 (in green on the same flux scale)\cite{fan03}, which is previously the most luminous high-redshift quasar known 
at $z=6.42$.\\
{\bf Figure 2. Transmission in absorption troughs and the proximity zone for J0100+2802}. Transmission in Ly$\alpha$ and Ly$\beta$ absorption troughs (shown in red and blue) were calculated by 
dividing the spectrum by a power-law continuum\cite{van01}, $F_{\lambda} \propto \lambda ^{-1.5}$. The shaded band shows 1 $\sigma$ standard deviation.   
The Ly$\alpha$ and Ly$\beta$ absorption redshifts are given by $\lambda$/$\lambda_{Ly\alpha(Ly\beta)}$-1, where $\lambda_{Ly\alpha}=1216\AA$ and 
$\lambda_{Ly\beta}=1026\AA$. The optical spectrum exhibits a deep Gunn-Peterson trough and a 
significant transmission peak at z $=$ 5.99. The proper proximity zone for J0100+2802 (in black) 
extends to $7.9\pm0.8$ Mpc, a much larger value than those of other $z>6.1$ quasars including $4.9\pm0.6$ Mpc for J1148+5251 (in green), 
consistent with its higher UV luminosity. The transmission in the bottom panel 
was calculated by dividing the measured spectrum 
by a power-law continuum $F_{\lambda} \propto \lambda ^{-1.5}$ plus two Gaussian fitting of 
Ly$\alpha$ and NV lines. The horizontal
dotted line and the two dashed lines denote transmission values of 0, 0.1 and 1.0
respectively, while the vertical dashed line denotes the proper proximity zone
size of 0.\\
{\bf Figure 3. The combined optical/near-infrared spectrum of J0100+2802 and the fitting of Mg II line}. Main panel, the black 
line shows the LBT optical spectrum and the red line shows the combined Magellan and Gemini near-infrared J,H,K-band spectra (from left to 
right, respectively). The gaps between J and H and between H and K bands are ignored due to the low sky transparency 
there. The magenta line shows the noise spectrum. The 
main emission lines Ly$\alpha$, C IV and Mg II are labeled. The details of the absorption lines are described in Extended Data Fig. 4. Inset,  fits of the Mg II line (with FWHM of $5,130\pm150~km~s^{-1}$) and 
surrounding Fe II emissions. The green, cyan, and blue solid lines show the power law (PL), Fe II and Mg II components. The black
dashed line shows the sum of these components in comparison with the observed spectrum, denoted by the red line. The black hole mass is estimated to be $(1.24\pm0.19)\times10^{10}M_{\odot}.$\\
{\bf Figure 4. Distribution of quasar bolometric luminosities, $L_{bol}$, and black hole masses, $M_{BH}$, estimated from the Mg II lines}. 
The red circle at top right represents J0100+2802. The small blue squares denote SDSS high redshift quasars\cite{jiang07,kurk07,derosa11}, and the large blue square represents J1148+5251. 
The green triangles denote CFHQS high-redshift quasars\cite{willott10,derosa11}. The purple star denotes 
ULAS J1120+0641 at $z=7.085$\cite{mortlock11}. Black contours (which indicate 1$\sigma$ to 5$\sigma$ significance from inner to outer) and grey dots denote SDSS low-redshift 
quasars\cite{shen11} (with broad absorption line quasars excluded). Error bars represent the 1$\sigma$ 
standard deviation, and the mean error bar for low redshift quasars is presented in the bottom-right corner.
The dashed lines denote the luminosity in different fraction of Eddington Luminosity, $L_{Edd}$. 
Note that the black hole mass and bolometric luminosity are calculated using 
the same method and the same cosmology model as in this paper, and the systematic uncertainties (not 
included in the erro bars) of virial black hole masses could be up to a factor of three\cite{peterson04}. 

%\newpage
\begin{figure}
\begin{center}
\leavevmode
\psfig{figure=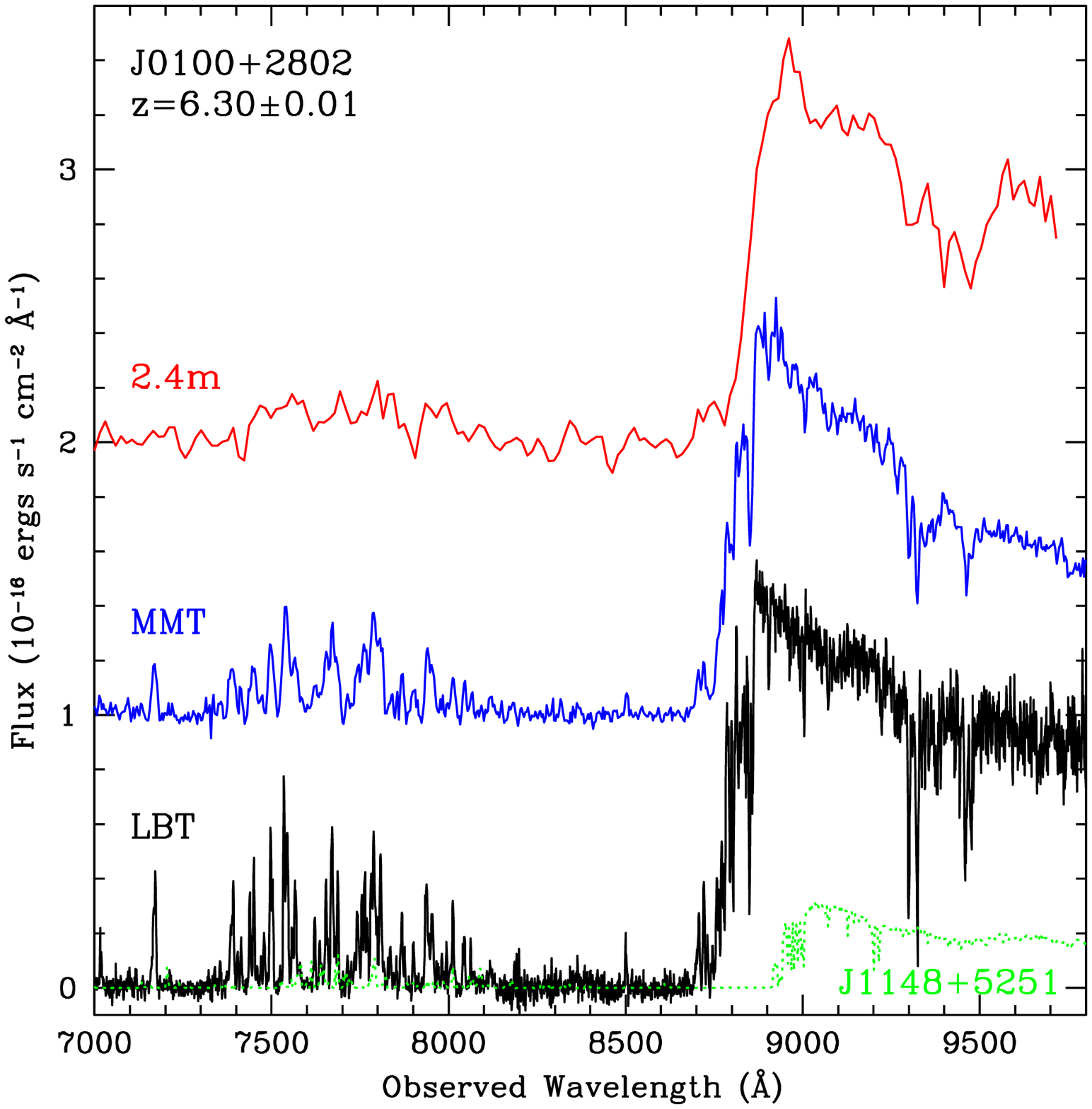,width=160truemm,angle=0}
\caption{} 
\label{}
\end{center}
\end{figure}
%\newpage
\begin{figure}
\begin{center}
%\leavevmode
\psfig{figure=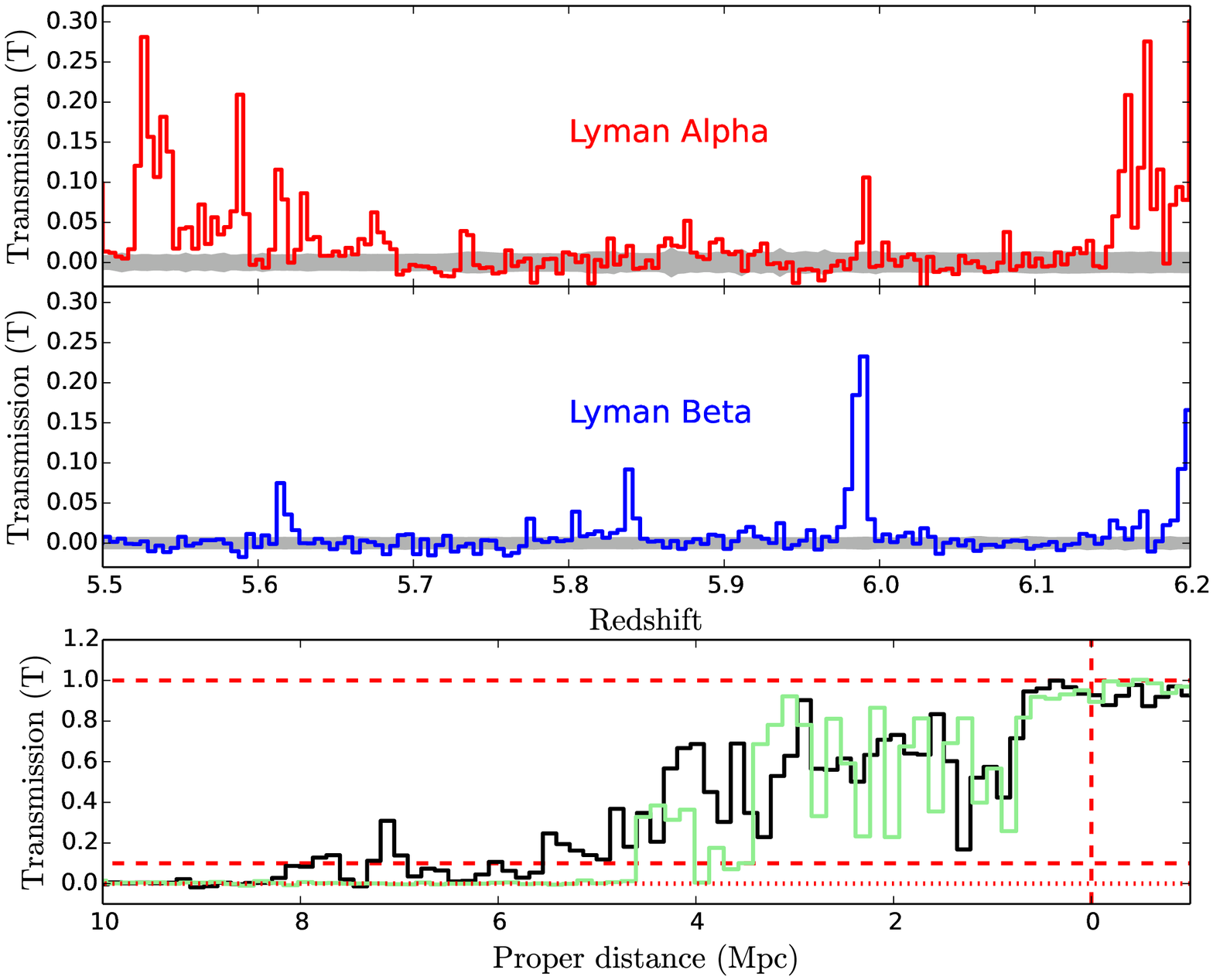,width=170truemm,angle=0}
\caption{}
\label{}
\end{center}
\end{figure}
%\newpage
\begin{figure}
\begin{center}
\leavevmode
\psfig{figure=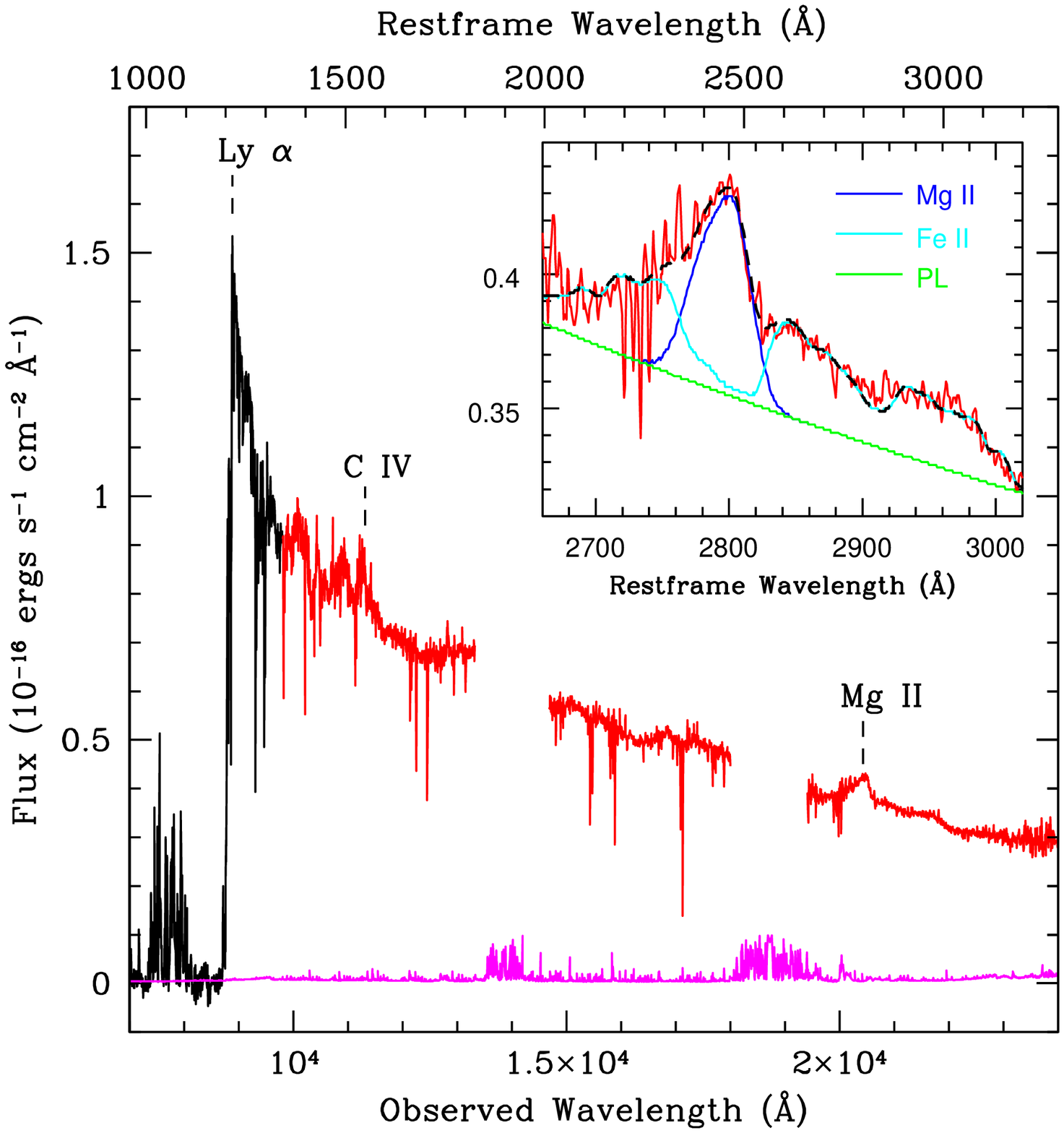,width=170truemm,angle=0}
\caption{}
\label{}
\end{center}
\end{figure}
%\newpage
\begin{figure}
\begin{center}
\leavevmode
\psfig{figure=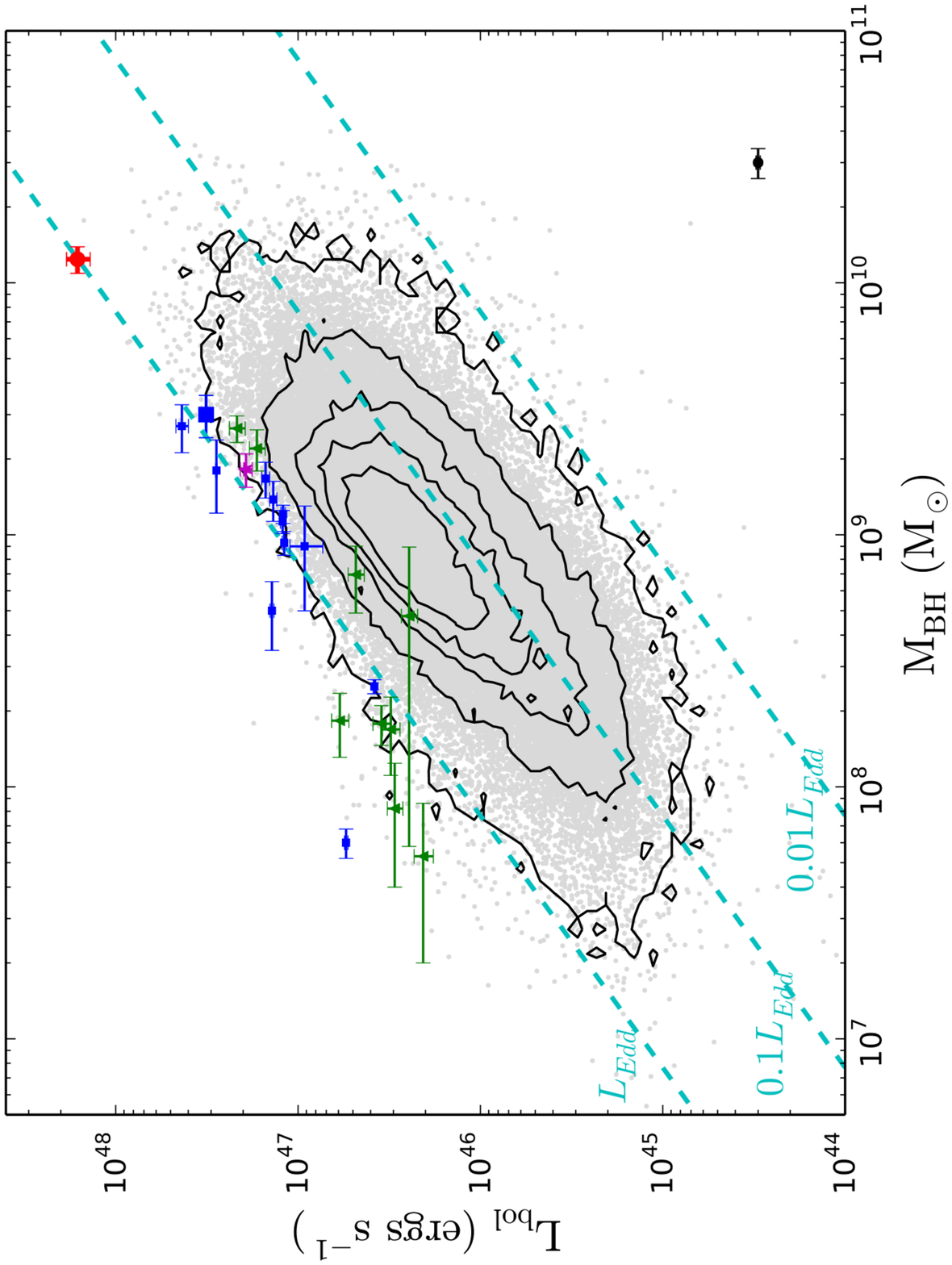,width=170truemm,angle=0}
\caption{}
\label{}
\end{center}
\end{figure}

\newpage

\noindent{\bf Methods}\\
The optical spectroscopy on J0100+2802 was first carried out on 29 December 2013 with 
the Yunnan Fainter Object Spectrograph and Camera (YFOSC) of the Lijiang 2.4-m telescope in China.
We used a very low resolution grism (G12, at a dispersion of 18$\AA$/pixel) and took 3,000s 
exposure on this target. The spectrum clearly shows a sharp break at about 8,800$\AA$ and no significant emissions at the blueward,
consistent with a quasar spectrum at redhift beyond 6.2. To confirm this discovery, two subsequent
optical spectroscopic observations were obtained on 9 and 24 January 2014 with the 6.5-m Multiple 
Mirror Telescope (MMT) and
the twin 8.4-m mirror Large Binocular Telescope (LBT) in the USA, respectively. The low to medium resolution
spectra, obtained with 1,200s exposure with MMT Red 
Channel (at a dispersion of 3.6$\AA$/pixel) and 2,400s exposure with LBT Multi-Object Double CCD Spectrographs/Imagers (MODS)\cite{pogge2010} 
(at a dispersion of 1.8$\AA$/pixel) respectively, explicitly confirm that SDSS J0100+2802 is a quasar 
at redshift of 6.30$\pm$0.01 (obtained by the $Ly \alpha$ line). 

The near-infrared K-band spectroscopy on J0100+2802 was carried out with LBT/LUCI-1 on 2 January 2014. Due to the short exposure time (15 minutes), the spectrum is of
modest signal-to-noise ratio (S/N). Although the Mg II line was clearly detected, the noisy LBT spectrum
did not allow us to accurately measure the line width. 
To improve the quality of near-infrared spectrum, we presented J,H,K-band spectroscopy with Gemini/GNIRS and Magellan/FIRE on 6 August and 7 October 2014, respectively. The exposure time was 3,600s for GNIRS and 3635s for FIRE. 
The FIRE spectrum has higher S/N (about 30 in K-band) and higher spectral resolution (R=$\lambda/\Delta\lambda\sim$ 6,000) than the GNIRS spectrum (with
S/N of about 10 in K-band and R$\sim$ 1,800). In order to achieve the best spectral quality, we combined both the FIRE 
and GNIRS spectrum, and 
scaled the combined spectrum according to its 2MASS J,H,$\rm{K_s}$-band magnitudes. The Mg II line shown in K-band spectrum
gives the same redshift as that given
by the $Ly \alpha$ line in the optical spectrum. The high-quality J,H,K-band spectra also clearly 
display abundant absorption
features, which have been identified to be from intervening or associated systems with redshifts from 2.33 to 6.14 
(see Extended Data Fig. 4).  

After redshift and Galactic extinction corrections, the rest-frame H and K-band spectrum is decomposed 
into a pseudo-continuum and the Mg II emission line.
The pseudo-continuum consists of a power-law continuum and Fe II emissions, and 
is fitted within the rest-frame wavelength range of between 2,000$\AA$ and 3,200$\AA$, by excluding
the boundary region between H and K-bands where the sky transparency is lower. An Fe II 
template\cite{vestergaard01,salviander07}
is adopted for the fitting of Fe II emissions.
The Mg II emission line is 
fitted with two broad Gaussian components. The four Mg II absorption lines near the redder part of Mg II emission line
are also fitted as 4 Gaussian lines in order to remove their effects to the fittings of Mg II and Fe II emission lines.
The overall FWHM (Full Width at Half Maximum) of Mg II emission line is  $\sim 5130~{\rm km~s}^{-1}$ 
with an uncertainty of 150~km~s$^{-1}$.  The continuum has a slope of -1.43 and the continuum 
luminosity at the rest-frame wavelength of 
3,000$\AA$ ~($L_{3,000}$) is $3.15\pm0.47 \times 10^{47} {\rm ergs~s}^{-1}$. 
The Fe II to Mg II line ratio is 2.56$\pm$0.18, which is consistent with the mean value of other z$>$6 quasars\cite{derosa11,derosa14}.
After applying a virial black hole
 mass estimator based on the Mg II line\cite{MD04}, we estimate its black hole mass to be $(1.24\pm 0.19) 
\times 10^{10} M_{\odot}$.  Although the systematic uncertainty of virial black hole mass estimation
can be up a factor of three\cite{peterson04},  our result still strongly indicates that J0100+2802 
hosts a central black hole with mass close to 10 billion  $M_{\odot}$. This is also well consistent with a black hole
mass obtained by assuming an Eddington lumonisity of J0100+2802, which
leads to a mass of $1.3\times 10^{10} M_\odot$.
Considering the contribution of Balmer continuum, as done for other z$>$6 quasars\cite{derosa11,derosa14}, leads to a decrease of 
$L_{3,000}$ to $2.90\pm0.44 \times 10^{47} {\rm ergs~s}^{-1}$, an increase of FWHM of Mg II to $5,300\pm200 ~{\rm km~s}^{-1}$ and yield a black-hole mass  of $(1.26\pm0.21) \times 10^{10}$ $M_\odot$. Therefore, the effect of considering Balmer
continuum is insignificant for the black hole mass measurement of J0100+2802. In addition, if we adopt a different
virial black hole mass scaling relation\cite{vo09}, the black hole mass will change to $(1.07\pm0.14) \times 10^{10}$ $M_\odot$, which is still well consistent with the result we obtained above.

\newpage
\noindent{\bf Legends for Extended Data Figures:}\\
{\bf Extended Data Figure 1.  Images of J0100+2802 in SDSS, 2MASS, and WISE bands}. J0100+2802 is undetected in SDSS u,g,r bands (top row) but is relatively bright in other bands (lower three rows). It is consistent with a point
source in the bands with high signal-to-noise detections. The size is 1$'$$\times$1$'$ for all images.  The green circle represents a angular size of 10$''$ in each image.\\
{\bf Extended Data Figure 2.  The  LBT K-band image of J0100+2802}. The size is 10$''$ $\times$10$''$. The horizontal and vertical axes denote the offsets in Right Asention ($\Delta$RA) and Declination ($\Delta$DEC).The 
image, with seeing of 0.4$^{\prime\prime}$, shows a morphology fully consistent with a point source. \\
{\bf Extended Data Figure 3. The rest-frame spectral energy distributions of J0100+2802, J1148+5251 and ULAS J1120+0641}. The redshifts of these three quasars are 6.30, 
6.42 and 7.085, respectively. The luminosity of J0100+2802 in the UV/optical bands is about 4 
times higher than that of J1148+5251, and
7 times higher than that of ULAS J1120+0641. The photometric data are from literature for J1148+5251 and J1120+0641. The error bars show the 1$\sigma$ standard deviation.\\
{\bf Extended Data Figure 4. The major absorption features identified from optical and near-infrared spectroscopy of J0100+2802}. Most of them are from Mg II, C IV and Fe II. The labels from A to H correspond to the redshifts of absorption materials at 6.14, 6.11, 5.32, 5.11, 4.52, 4.22, 3.34 and 2.33, respectively. Studies
of intervening and associated absorption systems will be discussed elsewhere.

%\newpage
\begin{figure}
\begin{center}
\leavevmode
\psfig{figure=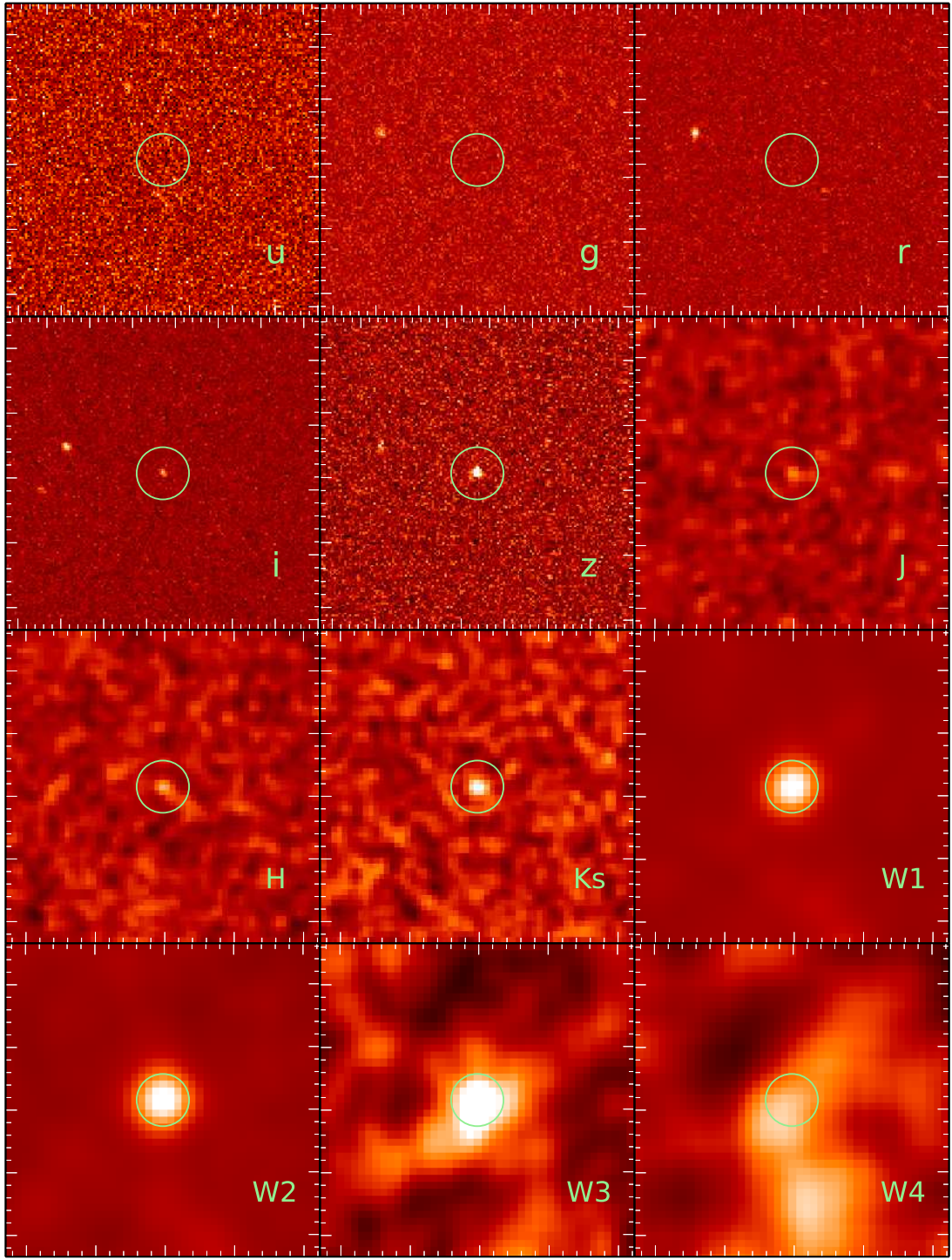,width=160truemm,angle=0}
\caption{(Extended Data Figure 1)} 
\label{}
\end{center}
\end{figure}
%\newpage
\begin{figure}
\begin{center}
%\leavevmode
\psfig{figure=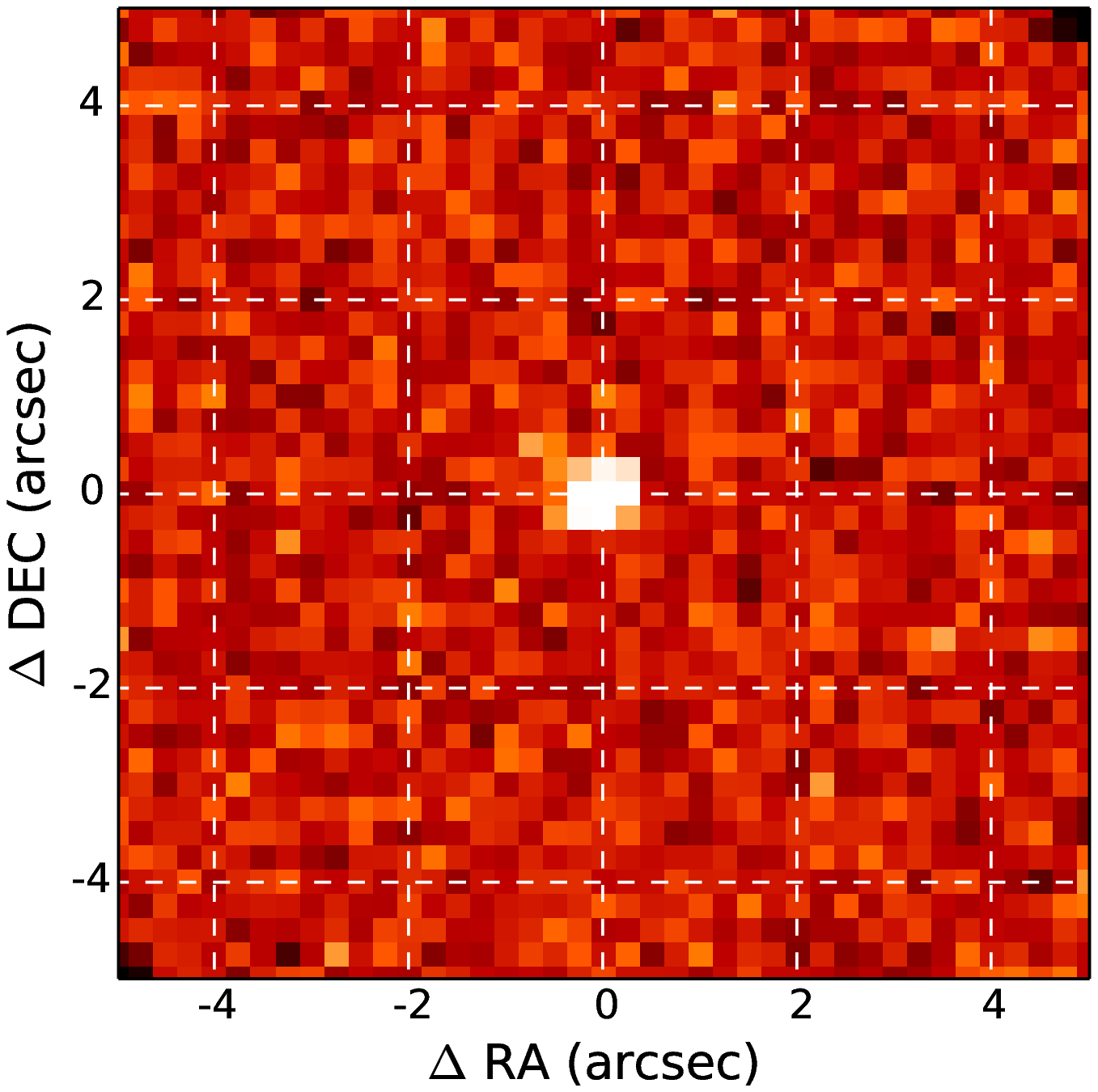,width=170truemm,angle=0}
\caption{(Extended Data Figure 2)}
\label{}
\end{center}
\end{figure}
%\newpage
\begin{figure}
\begin{center}
\leavevmode
\psfig{figure=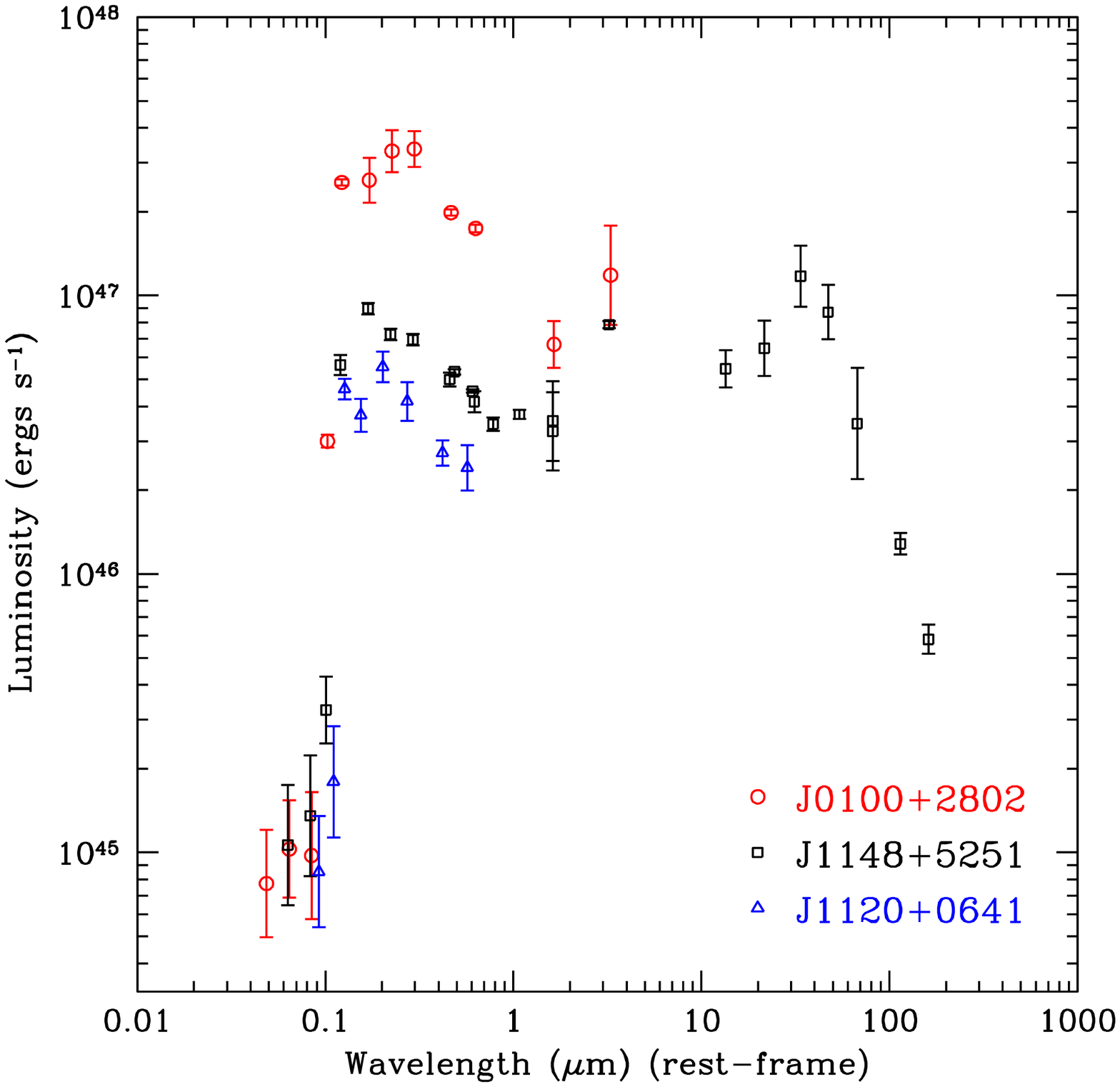,width=170truemm,angle=0}
\caption{(Extended Data Figure 3)}
\label{}
\end{center}
\end{figure}
%\newpage
\begin{figure}
\begin{center}
\leavevmode
\psfig{figure=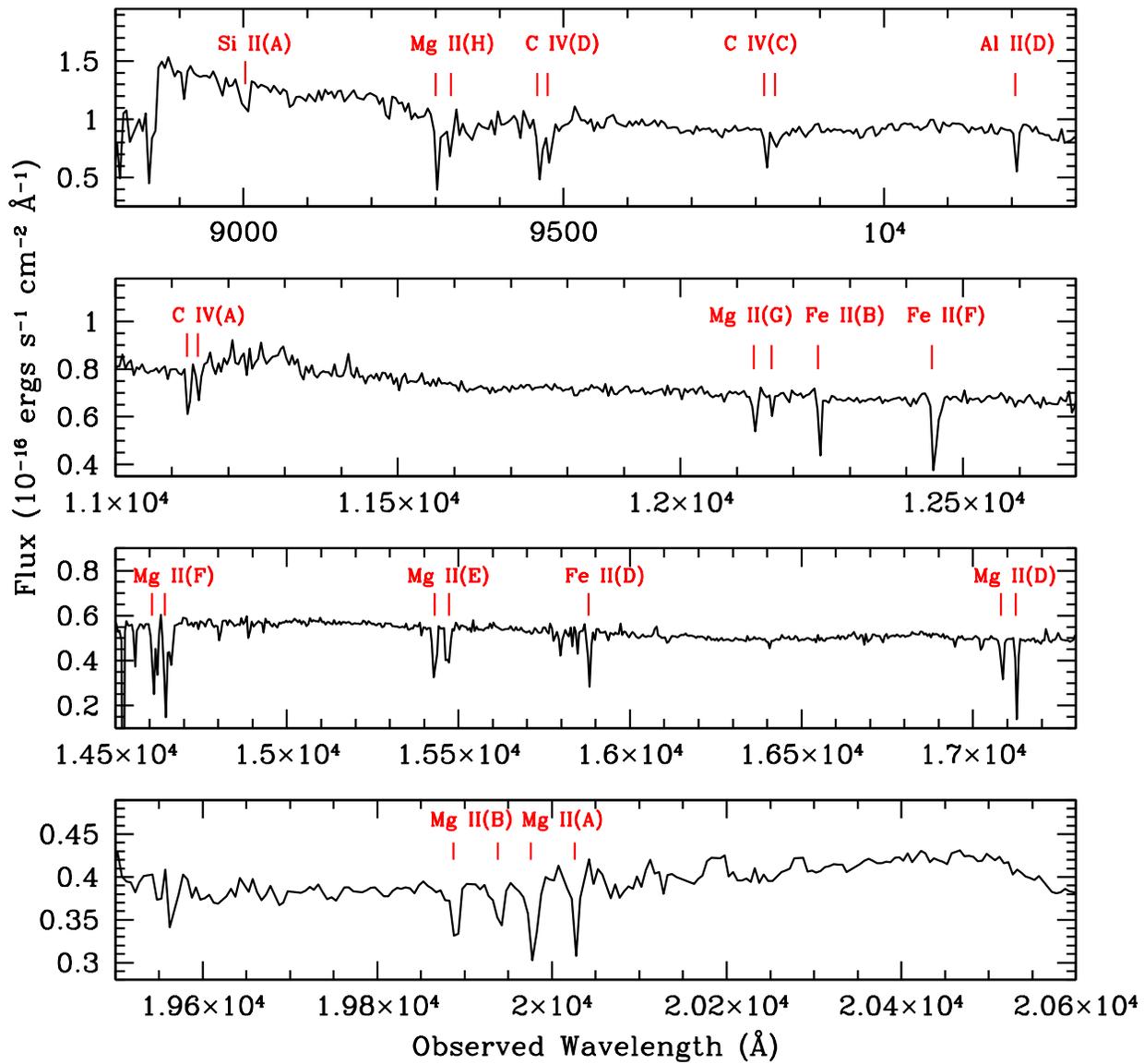,width=170truemm,angle=0}
\caption{(Extended Data Figure 4)}
\label{}
\end{center}
\end{figure}

\newpage

\noindent{\bf Additional Acknowledgments:}\\
Observations reported here were obtained at the MMT Observatory, a joint facility 
of the University of Arizona and the Smithsonian Institution. This work is based on data 
acquired using the Large Binocular Telescope (LBT). The LBT is an international 
collaboration among institutions in the United States, Italy and Germany. LBT Corporation 
partners are: The University of Arizona on behalf of the Arizona university system; 
Istituto Nazionale di Astrofisica, Italy; LBT Beteiligungsgesellschaft, Germany, representing the
Max-Planck Society, the Astrophysical Institute Potsdam, and Heidelberg University; The Ohio State 
University, and The Research Corporation, on behalf of The University of Notre Dame, University of Minnesota and
University of Virginia. This paper uses data taken with the MODS spectrographs built with funding from NSF grant AST-9987045 and the 
   NSF Telescope System Instrumentation Program (TSIP), with additional funds from the Ohio Board of Regents and 
   the Ohio State University Office of Research. 
This paper includes data gathered with the 6.5 meter Magellan Telescopes located at Las Campanas Observatory, Chile.
This work is based on observations obtained at the Gemini Observatory, which is operated by the Association of 
Universities for Research in Astronomy, Inc., under a cooperative agreement with 
the NSF on behalf of the Gemini partnership: the National Science Foundation 
(United States), the National Research Council (Canada), CONICYT (Chile), the 
Australian Research Council (Australia), MinistÈrio da CiÍncia, Tecnologia e 
InovaÁ„o (Brazil) and Ministerio de Ciencia, TecnologÌa e InnovaciÛn Productiva 
(Argentina).
Funding for SDSS-III has been provided by the Alfred P. Sloan Foundation, the Participating Institutions, the National Science Foundation, and the U.S. Department of Energy Office of Science. The SDSS-III web site is http://www.sdss3.org/.
SDSS-III is managed by the Astrophysical Research Consortium for the Participating Institutions of the SDSS-III Collaboration including the University of Arizona, the Brazilian Participation Group, Brookhaven National Laboratory, Carnegie Mellon University, University of Florida, the French Participation Group, the German Participation Group, Harvard University, the Instituto de Astrofisica de Canarias, the Michigan State/Notre Dame/JINA Participation Group, Johns Hopkins University, Lawrence Berkeley National Laboratory, Max Planck Institute for Astrophysics, Max Planck Institute for Extraterrestrial Physics, New Mexico State University, New York University, Ohio State University, Pennsylvania State University, University of Portsmouth, Princeton University, the Spanish Participation Group, University of Tokyo, University of Utah, Vanderbilt University, University of Virginia, University of Washington, and Yale University. 
This publication makes use of data products from the Two Micron All Sky Survey, which is a joint project of the University of Massachusetts and the Infrared Processing and Analysis Center/California Institute of Technology, funded by the National Aeronautics and Space Administration and the National Science Foundation.
This publication makes use of data products from the Wide-field Infrared Survey Explorer, which is a joint project of the University of California, Los Angeles, and the Jet Propulsion Laboratory/California Institute of Technology, funded by the National Aeronautics and Space Administration.

\end{document}